\documentclass[12pt,preprint]{aastex}
\usepackage{graphicx}
\usepackage{subfigure}
\usepackage{subfloat}
\usepackage{longtable}
\usepackage{multirow}
\usepackage{color}
\usepackage{url}
\usepackage{lipsum}
\usepackage{booktabs}

\begin{document}
\title{The Nuclear Activities of Nearby S0 Galaxies}
\author{Meng-Yuan Xiao\altaffilmark{1,2,3}, Qiu-Sheng Gu\altaffilmark{1,2,3}, Yan-Mei Chen\altaffilmark{1,2,3} and Luwenjia Zhou\altaffilmark{1,2,3}}
\affil{$^1$ School of Astronomy and Space Science, Nanjing University, Nanjing 210093, P.~R.~China; \email{qsgu@nju.edu.cn}}
\affil{$^2$ Key Laboratory of Modern Astronomy and Astrophysics, Nanjing University, Nanjing 210093, P.~R.~China}
\affil{$^3$ Collaborative Innovation Center of Modern Astronomy and Space Exploration, Nanjing 210093, P.~R.~China}

\begin{abstract}

We present a study of nuclear activities in nearby S0 galaxies. After cross-matching the Sloan Digital Sky Survey Data Release 7 (SDSS DR7) with the Third Reference Catalog of Bright Galaxies (RC3) and visually checking the SDSS images, we derive a sample of 583 S0 galaxies with the central spectrophotometric information. In order to separate nebular emission lines from the underlying stellar contribution, we fit the stellar population model to the SDSS spectra of these S0 galaxies. According to the BPT diagram, we find that $8\%$ of S0 galaxies show central star-forming activity, while the fractions of Seyfert, Composite and LINERs are 2\%, 8\%, and 21.4\%, respectively. We also find that  star-forming S0s have the lowest stellar masses, over one magnitude lower than the others, and that the active S0s locate mainly in the sparse environment, while the normal S0s in the dense environment, which might suggest that the environment plays an important role in quenching star formation and/or AGN activity in S0 galaxies. By performing bulge-disk decomposition of 45 star-forming S0s in {\em g}- and {\em r}-bands with the 2D fitting software \textsc{Galfit}, 
 as well as exploiting the catalog of 2D photometric decompositions of \citet{Meert2015},
 \textbf{we find that the bulges of $\sim$ 1/3 star-forming S0 galaxies (16/45) are bluer than their disks}, while for other types of S0s, the bulge and disk components show similar color distributions. Besides, the S\'{e}rsic index of most star-forming S0s bulges is less than 2, while for normal S0s, it is between 2 and 6.

\end{abstract}


\section{Introduction}
In the Hubble `tuning fork' galaxy morphological classification scheme \citep{Hubble1936},  lenticular (S0) galaxies lie at an intermediate position between ellipticals and spirals. Such S0 galaxies, not only have disky morphology like spiral galaxies, but have red color and large bulges like elliptical galaxies, as well implying that they contain mainly old stellar population.

The formation and evolution of S0 galaxies are very important for understanding the formation and evolution of galaxies, but they are still an open question \citep[e.g., the recent review by][]{D'Onofrio2015}. 
Currently there are two possible scenarios  on the origin of S0 galaxies. One is that S0 galaxies are transformed from spiral galaxies, where spirals lose their gas and star formation is rapidly quenched. The other is that S0 galaxies are intrinsically different from spiral galaxies since their formation \citep{Kormendy2004,Barway2009, van2009}. The transformation origin may be associated with intra-cluster medium and neighboring galaxies, via minor mergers, slow encounters, galaxy harassments \citep{Moore1996} or tidal effects in dense environment \citep{Gunn&Gott1972, Larson1980, Dressler1983, Mihos1994, Moore1998, Moore1999, Neistein1999,Shioya2002}. A lot of studies have discussed the environmental dependence of galaxy evolution. Generally speaking, early-type galaxies tend to be in dense environment and have low star formation rate (SFR) \citep{Dressler1980,Balogh1997,Balogh1998,Balogh2000,Poggianti1999,Treu2003}. The fraction of S0 galaxies in the field is only about 15 \%, while spirals are the majority \citep{Naim1995}. Within the group environment, spirals and S0s are both about 40-45 \% \citep{Postman1984}. Furthermore, S0s become dominant in dense environment, the fraction grows up to 60 \% in clusters \citep{Dressler1980,Postman1984}. \citet{Dressler1997}, \citet{Fasano2000} and \citet{Desai2007} also found that the galaxy morphological distributions change abruptly in clusters at $z=0.4 \sim 0.5$,  about $50 \sim 70$ \% of spirals at high redshift ($z$ $>$ 0.4) are transformed into S0s, while the fraction of ellipticals, about 25 \%, remains nearly constant between $z=0.8$ and $z=0.0$. 
However, \citet{Wilman2009} showed that the fraction of S0s in groups is the same as in clusters but  it is much higher than in the field at $z=0.4$, which might suggest that S0s are formed in groups or subgroups. 

The Sloan Digital Sky Survey (SDSS) is one of the most ambitious and successful astronomical survey projects, both photometric and spectroscopic, ever undertaken \citep{Gunn1998,Blanton2003}, which uniformly surveyed more than  a quarter of the entire sky and provided photometric ($u$, $g$, $r$, $i$, and $z$ bands) and spectroscopic data for nearly one million galaxies. 
By using SDSS, \citet{Salim2012} found that in the $UV$-$optical$ color-stellar mass diagram (CMD), about half of S0 galaxies are located in the green-valley region, suggesting that these S0 galaxies might be transforming from actively star-forming to passive galaxies. \citet{Schawinski2007a} found that there are about 4\% among $\sim$16,000 early-type galaxies showing ongoing star formation. This percentage can be even as high as 30\% by using near-ultraviolet data from GALEX \citep{Kaviraj2007}. Recent studies indicate that the fraction of star-forming early-type galaxies depends on observational wavelengths and galaxy masses \citep{Schawinski2006,Schawinski2007b}.  \citet{cv06} showed that low-ionization nuclear emission-line regions (LINERs) appear to be even more popular in S0 galaxies \citep{Magliocchetti1999,Magliocchetti2004,Overzier2003}.

We have two main purposes in this work, one is to obtain the fundamental physical properties of nearby S0 galaxies, such as stellar mass, velocity dispersion, {\em r}-band absolute magnitude, and  {\em g - r} color of different types of active S0 galaxies, and the other is to study the relationship between the central activity of S0 galaxies and their environments. 

This paper is organized as it follows. The sample selection and data reduction are shown in Section 2. In Section 3, we present the analysis and the main results, including the star formation rates and some basic physical properties. We concentrate on interpreting the dependence between S0 activity and environment. We also perform the bulge-disk decompositions to compare bulge and disk colors. In Section 4 we report our conclusions. We assume a flat $\Lambda$CDM cosmology with $\Omega_m=0.3, \Omega_\Lambda=0.7$ and $H_0=70$ $km^{-1}$ $s^{-1}$ $Mpc^{-1}$ in this paper.

\section[]{The Data}
\subsection{Sample Selection}
The Third Reference Catalog of Bright Galaxies\footnote{http://haroldcorwin.net/rc3/} \citep[RC3;][]{deVauc91} is one of the most popular catalog of nearby, large, and bright galaxies, which contains 23,011 galaxies with 
the apparent major isophotal diameter, $D_{25}$, larger than one arcminute, B-band magnitude brighter than $15.5 \ mag$, and the redshift lower than 15,000 km s$^{-1}$. 
Recently, \citet{Lee2015} implemented a recursive algorithm that determined the correct position for each galaxy automatically and updated the RC3 catalog appropriately. We start with this version of RC3 catalog and pick out all galaxies with the revised morphological classification \textbf{of Lenticulars. We find that there} are 3472 lenticular (also S0) galaxies in RC3.


We cross-match these 3472 S0 galaxies with SDSS Data Release 7 spectroscopic archive database\footnote{http://www.sdss.org} \citep[DR7;][]{Abazajian2009}, whose spectra are obtained with 3-arcsec fibers, cover a wavelength range of 3800-9200 \AA, and have a spectral resolution of 1800-2200. By constraining the position-matching accuracy to 3-arcsec,
we obtain a sample of 694 S0 galaxies. 

In order to be sure that our sample does not contain any spiral galaxy, we check the morphological classifications of these 694 galaxies in Galaxy Zoo 2\footnote{http:/zoo2.galaxyzoo.org/} \citep[GZ2;][]{Willett2013}, which provides the morphological features of 304,122 galaxies from SDSS DR7 with $m_r < 17 \ mag$, including bars, bulges, disk shapes, spiral arms, etc. We rule out 92 galaxies having GZ2 class $t$, $m$, $l$, $1$, $2$, $3$, $4$, $+$, which mean the sign of spiral arm pattern and the number of spiral arms. Furthermore, by visually checking SDSS \textbf{{\em g-}, {\em r-}, and {\em i-}band} images one by one, we remove other 19 galaxies with apparent spiral structure. Finally, we get a sample of 583 S0 galaxies with highly reliable morphological classification, as well as photometric ({\em $u$, $g$, $r$, $i$, and $z$}) and spectrophotometric data from SDSS. The median spectral signal-to-noise ratio (S/N) per pixel of the sample is $\sim$ 45.

\subsection{Data Analysis}
In order to separate nebular emissions from the stellar continuum and absorptions, we closely follow the method discribed in \citet{Tremonti2004} and \citet{Brinchmann2004} to fit stellar population models to the continuum of each galaxy spectrum. The single stellar population models are generated using a preliminary version of the population synthesis code of Charlot \& Bruzual (2016, in preparation) which incorporates the MILES empirical stellar library \citep{Sanchez2006}. In this work we make use of stellar populations with 10 different ages (0.005, 0.025, 0.1, 0.2, 0.6, 0.9, 1.4, 2.5, 5, and 10 Gyr) and four metallicities (0.004, 0.008, 0.017, and 0.05). We first mask out the strong nebular emission lines. For each metallicity, the 10 template spectra are convolved with the measured stellar velocity dispersion (from SDSS spectrophotometric catalog) of each individual galaxy, and the best fit model spectrum is expressed as a non-negative linear combination of the 10 template spectra with dust extinction \citep{Charlot2000} modeled as an additional free parameter. The best fit model is then expressed as a linear combination of single stellar population models under the basic assumption that the star formation history of any galaxy can be approximated by a sum of discrete bursts. The metallicity yielding the minimum $\chi^2$ is selected as the final best fit. The top panel of Fig. 1 shows an example of continuum fitting, the black line is the observed spectrum and the red line is the best fit model. By subtracting the best-fit model, we get the pure emission line spectrum (the black line spectrum in the bottom panels), each emission line is then fitted with one Gaussian component (green line). For the H$\beta$ region, we fit H$\beta$, [{O~\sc iii}]$\lambda$4959 and [{O~\sc iii}]$\lambda$5007 lines together. The line centers and widths of H$\beta$ and [{O~\sc iii}]$\lambda$4959 are tied to those of [{O~\sc iii}]$\lambda$5007. For the H$\alpha$ region, we fit H$\alpha$, [{N~\sc ii}]$\lambda\lambda$6548,6584 lines together. The line centers and line widths of [{N~\sc ii}]$\lambda\lambda$6548,6584 are tied to those of H$\alpha$.

\section[]{Analysis and Results}
\subsection{Classification of the S0 sample}
\citet[thereafter BPT]{Baldwin1981} suggested that it was possible to separate type 2 AGNs from star-forming galaxies using the intensity ratios of two pairs of relatively strong emission lines, close in wavelengths to avoid reddening effects (e.g. [{N~\sc ii}]$\lambda$6584 /H$\alpha$, [{O~\sc iii}]$\lambda$5007 /H$\beta$). This technique was then refined by lots of following works, i.e., \citet{Veilleux1987}, \citet{Kewley2001}, \citet{Kauffmann2003a}. Following the criteria of \citet{Kewley2001} and \citet{Kauffmann2003a}, we use the BPT diagnostic diagram to discriminate between different nuclear activities of nearby S0 galaxies according to their different ionization mechanisms estimated from the emission line flux ratios, [{N~\sc ii}]$\lambda$6584 /H$\alpha$ and [{O~\sc iii}]$\lambda$5007 /H$\beta$. In this work, we first classify the S0 galaxies according to the S/N of the four emission lines. We refer to the galaxies as active S0s, that is S0s with obvious star-formation or/and AGN activity, when S/N $>$ 5 for the four emission lines, and as Others in case of S0s with very low-level emission lines, that is S/N $<$ 3, or quiescent S0s with classic absorption line spectra. In our sample, there are 232 S0s have S/N $>$ 5 for the four emission lines. By means of the diagnostic diagram, we classify these 232 active S0 galaxies into four sub-samples according to their emission line ratios: 45 star-forming galaxies, 49 composite galaxies, 13 Seyfert galaxies, and 125 LINERs. Fig. 2 shows the distribution of 232 S0 galaxies on the BPT diagram, the red solid/dashed lines correspond to the AGN/star forming galaxies separation lines of \citet{Kewley2001}  and \citet{Kauffmann2003a}, respectively. Seyfert galaxies are often defined to have $\log$([{O~\sc iii}]/H$\beta$) $>$ 3 (horizontal blue line), and LINERs to have $\log$([{O~\sc iii}]/H$\beta$) $<$ 3. Fig. 3 shows examples of SDSS false-color images (left) with the corresponding spectral fitting (right) of Star-forming, Composite, Seyfert, LINERs, and Others S0 galaxies classified in the present study. The observed spectra are shown in black and the best fit models as the red color in right panels.

\subsection{Galaxy Properties}

The physical parameters we adopt in this study include stellar mass for the whole galaxy ($\rm M_\ast$), stellar velocity dispersion ($\sigma_\ast$), redshift ($z$), SDSS {\em r}-band absolute magnitude ($\rm M_r$), and SDSS $g-r$ color. The stellar mass  ($\rm M_\ast$) is extracted from MPA/JHU catalog\footnote{http://wwwmpa.mpa-garching.mpg.de/SDSS/DR7/} which is based on SDSS {\em $u$}{\em $g$}{\em $r$}{\em $i$}{\em $z$} photometry and the methodology detailed in \citet{Brinchmann2004} and \citet{Salim2007}. The $\sigma_\ast$ is computed from fitting the absorption lines in the SDSS spectra after masking emission-line regions and the redshift is derived by fitting the emission lines in the SDSS spectra. The $\rm M_r$ is taken from the New York University Value Added Galactic Catalog \citep[NYU-VAGC;][]{Blanton2005}\footnote{http://sdss.physics.nyu.edu/vagc/}. We adopt the {\em r}-band Petrosian magnitude with k-correction. Table 1 summarizes the median values and scatters of galaxy properties for different S0s, including the number of different types of S0 galaxies, stellar mass, velocity dispersion, the area corresponding to SDSS 3-arcsec fiber, $z$, $\rm M_r$ and $g-r$ color. The values in parentheses represent the $\pm$ 1$\sigma$ error. Fig. 4 shows the histograms of different types of S0s as a function of stellar mass (top-left), velocity dispersion (top-right), redshift (bottom-left) and {\em r}-band absolute magnitude (bottom-right), respectively. The vertical lines show median values of the physical parameters for different types of S0s. The black filled columns represent star-forming S0s, the orange, blue, green, and red histograms show Composite, Seyfert, LINERs, and Others, respectively. From Table 1 and Fig. 4, it is clear that the physical parameters of S0s which are classified as Composite, Seyfert, LINERs and Others are similar. The only exception is the star-forming S0s (hereafter SFS0s), which are more nearby, the stellar mass is over one order of magnitude lower than other types of S0s. 



Fig. 5 shows the correlation between different physical parameters for S0 galaxies, including (1) stellar mass versus velocity dispersion ($M_\ast$ vs. $\sigma_\ast$, top-left), (2) stellar mass versus {\em r}-band absolute magnitude ($M_\ast$ vs. $M_r$, top-right), (3) stellar mass versus the $g-r$ color ($M_\ast$ vs. $g-r$, bottom-left), and (4) {\em r}-band absolute magnitude versus the $g-r$ color ($M_r$ vs. $g-r$, bottom-right).The black, orange, blue, green, and red points represent SFS0s, Composite, Seyfert, LINERs, and Others, respectively. The background grey contours show 135,912 galaxies from MPA/JHU catalog with median S/N per pixel of the spectra larger than 20. We find that our S0 galaxies are in agreement with the distribution of SDSS galaxy sample quite well. The linear (Pearson) correlation coefficient and the p-value of different physical properties for different types of S0s are listed in Table 2. We use the Kolmogorov-Smirnov (K-S) analysis to compare different properties of the five types of S0s. Values of K-S statistical probability of $M_\ast$, $M_r$ and $g-r$ color for SFS0s compared to Composite, Seyfert, LINERs, and Others from the same parent population are listed in Table 3. The results suggest that SFS0s are significantly different from other types of S0 galaxies. The tendency of SFS0s to be less massive, bluer and fainter than other types of S0s might be explained by the `down-sizing' effect. \citet{Cowie1996} suggested that the most massive galaxies were formed earliest in the Universe (i.e. at high redshift), and star-forming activity is progressively moved from massive galaxies to smaller, lower-mass ones, although their data are limited to galaxies brighter than $L^*$ (characteristic luminosity, roughly comparable in luminosity to our Milky Way). This might suggest that more massive S0s evolve faster than their less massive counterparts. 

\subsection{Nuclear star formation}
For 45 star-forming S0 galaxies, we calculate the central 3-arcsec SFR from the ${H_\alpha}$ luminosity \citep{Kennicutt1998}:
    \begin{equation}
        {\rm SFR}~(M_\odot~yr^{-1}) = 7.9 \times 10^{-42}~L(H_\alpha)~({\rm ergs~s^{-1}}).
    \end{equation}

\noindent Here the ${H_\alpha}$ emission has been corrected for nebular extinction, $A_{V,nebular}$ \citep{Cardelli1989}, which is derived from the Balmer decrement $F_{H_\alpha}/F_{H_\beta}$:
 \begin{equation}
       A_{V,nebular} = 7.2 \times\ \log\
       (\frac{F_{H_\alpha}/F_{H_\beta}}{I_{H_\alpha}/I_{H_\beta}}),
   \end{equation}
\noindent where $F_{H_\alpha}/F_{H_\beta}$ and $I_{H_\alpha}/I_{H_\beta}$ are the observed and intrinsic flux ratios of H$_\alpha$ and H$_\beta$, respectively, we adopt $I_{H_\alpha}/I_{H_\beta}$ = 2.86 \citep{Osterbrock1989}. The distribution of $A_{V, nebular}$ and SFRs are shown in Fig. 6. We find that the mean values of nebular extinction, $log$SFR and $log$SFR per unit area  within the SDSS fiber size are 1.06 $\pm$ 0.66,  $-$1.13 $\pm$ 1.07 $M_{\odot}$ $yr^{-1}$, and $-$0.48 $\pm$ 0.56 $M_{\odot}$ yr$^{-1}$ kpc$^{-2}$, respectively.

In order to compare the nuclear star formation activities of 45 star-forming S0 galaxies to those of star-forming non-S0 galaxies, we construct a reference sample of star-forming non-S0 galaxies from the MPA/JHU DR7 catalog on the basis of two parameters: the stellar mass and the redshift. The baseline tolerances used for matching are 0.1 dex in $log M_\ast$ and 0.001 in $z$. The motivation for choosing this set of matching parameters are the following: (1) constraining the reference galaxies with similar $M_\ast$ because the stellar population properties vary strongly as a function of the stellar mass; (2) constraining the reference galaxies with similar $z$ in order to avoid the effect of evolution. For each SFS0 galaxy, the SFR of the reference sample is taken as the median value of the matched star-forming non-S0 galaxies. The mean nuclear $log$SFR, calculated from the ${H_\alpha}$ luminosity, is $-$1.75 $\pm$ 1.00 $M_{\odot}$ $yr^{-1}$ and mean $log$SFR per unit area is $-$1.10 $\pm$ 0.46 $M_{\odot}$ yr$^{-1}$ kpc$^{-2}$, which suggest that the star-forming activities (both SFR and SFR per unit area) in SFS0s and non-S0 galaxies are similar within the errors.



\subsection{Bulge and Disk Decomposition}

Since the SDSS spectra only provide us with the physical information of the central regions of S0 galaxies, in order to know the behavior of whole galaxies, we make use of SDSS {\em g}- and {\em r}-band images. Fortunately, \citet{Meert2015} presented a catalog of two-dimensional bulge-disk photometric decomposition of $\sim7\times10^{5}$ spectroscopically selected galaxies of SDSS DR7. These galaxy images are fitted with two-dimensional profiles, PSF-corrected de Vacouleurs (\texttt{Dev}), S\'{e}rsic (\texttt{Ser}), de Vacouleurs+Exponential (\texttt{DevExp}), S\'{e}rsic+Exponential (\texttt{SerExp}) profiles and performed for the SDSS {\em g}-, {\em r}- and {\em i}-bands utilizing the fitting routine \textsc{Galfit} and analysis pipeline \textsc{PyMorph}. We choose the \texttt{SerExp} catalog which uses the S\'{e}rsic model for bulges and the Exponential model for disks (Meert, private communication). Meert's catalog includes 670,722 galaxies with the redshift $>$0.005 and the extinction-corrected $r$-band Petrosian magnitude between 14 and 17.77 mag, since the 17.77 mag in $r$-band is the lower magnitude-limit for completeness of the SDSS main spectroscopic galaxy sample, and the limit of 14 mag is to exclude large, nearby galaxies which are too well resolved to be fitted with the standard smooth light profile. Even if most of our S0s are very nearby galaxies, there are only 154 S0s contained in the Meert's catalog. We select galaxies with reliable decomposition from Meert's catalog using the following criteria: (1) the flag bits from $1$ to $13$, (2) the ``nbulge'' less than 8, (3) the ``numtargets'' no larger than 1, (4) the ``numneighborfits'' no larger than 0, (5) without unusual values (such as ``mbulge'' and ``mdisk'' set to 999). The first criterion ensures that we selected galaxies with a reasonable fit. The second criterion is set for obtaining reliable sub-components (bulge and disk) magnitude. The last three criteria are applied to minimize contamination from the redundant sources within 3-arcsec of the centre of the image, to avoid neighbour sources fitted simultaneously to the target galaxy, and to exclude unusual values. Considering all the above limitation, we finally get only 76 S0s in our sample.


Among these 76 S0 galaxies, there are only 9 SFS0s. In order to improve the statistics, we perform decompositions of 45 SFS0s in our sample in {\em g-} and {\em r-} bands by utilizing the \textsc{Galfit} \citep[version 3.0.5;][]{Peng2002,Peng2010} software. \textsc{Galfit} is a tool for extracting information about galaxies by using parametric functions to fit models to light profiles in two-dimensional digital images. It adopts the Levenberg-Marquardt algorithm based on the least-square minimization to find the optimum solution to fit. Because the disks of S0 galaxies are not usually pure exponential disks, we adopt two S\'{e}rsic light profiles for the bulge and the disk, respectively \citep{Sersic1968}. \textbf{Even if the seeing is not very good (the median value for the $r$-band is $1.43''$), our S0s are nearby and we let the S\'{e}rsic indexes $n$ free to vary in order to characterize the bulge and disk types.} The S\'{e}rsic light profile used extensively to study galaxy morphology is expressed as follows:
\begin{eqnarray}
I(R) = I_{\rm e}\  {\rm exp}\left\{-b_{\rm n}[(R/R_{\rm e})^{1/n} -1]\right\},
\label{sersic-eq}
\end{eqnarray}
\begin{eqnarray}
b_{n} \simeq 1.9992n - 0.3271.
\label{bn1}
\end{eqnarray}
\noindent Where R is the distance from the center of a galaxy, $I(\rm R)$ is the surface brightness at  $R$, $I_{\rm e}$ is the surface brightness at the effective radius $R_{\rm e}$ which is known as that half of the total light of a galaxy within  $R_{\rm e}$, and the S\'{e}rsic index $n$ shapes the light profile of a galaxy.


Input image (observed image), $\sigma$ image (background noise image), PSF image and bad pixel mask image are four primary data used in the decomposition. The input and PSF images of 40 S0s are drawn from the NASA-Sloan Atlas\footnote{http://www.nsatlas.org/data}, for other 5 S0s left out of NASA-Sloan Atlas, the input images are taken from SDSS DR12 Science Archive Server\footnote{http://data.sdss3.org/fields/}, while the PSF images are reconstructed from the psField files\footnote{https://www.sdss3.org/dr8/imaging/images.php}. The $\sigma$ images are internally calculated by \textsc{Galfit} after setting the image header keywords: EXPTIME (exposure time), GAIN, and NCOMBINE (the number of combined fits). The bad pixel mask images are created manually using \textsc{SAOImage DS9} \citep{Joye2003} regions. When the neighboring galaxies or foreground stars are too close causing the contamination of the target galaxy, we add two S\'{e}rsic models and one PSF model to fit one neighboring galaxy and one foreground star instead of masking them, respectively.
  
Before runnig \textsc{Galfit}, a list of parameters must be set. One part is the image parameters and the other is the object fitting parameters. The image parameters containing four primary images and other parameters such as image region to fit, size of the convolution box, and magnitude photometric zero point can be easily set and are fixed with fitting. \textsc{Galfit} takes a simultaneous fitting of numbers of model components with the the object fitting parameters setted for each component. We enter the initial rough guesses based on the SDSS observation quantities (such as $r\_deV$, $r\_exp$, $ab\_deV$, $ab\_exp$, $phi\_deV$, $phi\_exp$, $deVMag$, and $expMag$) and these parameters are changed by \textsc{Galfit} to achieve normalized chi-squared ($\chi^{2}$) close to 1 with low residuals.
  
We run \textsc{Galfit} to 45 SFS0s, the mean values of normalized $\chi^{2}$ in {\em g}- and {\em r}-bands are 1.05 and 1.04, respectively. We show three examples of  \textsc{Galfit}  decomposition in Fig. 7. In the top panels the $g$-band image, the model image, and the residual image of PGC 36750 are displayed (from left to right). The middle and bottom panels show the results for the $r$-band images of PGC 49734 and PGC 38112.

In order to test the reliability of our bulge/disk decompositions, we compare our results with Meert's ones for the common 9 SFS0s, and we find that for 8 SFS0s, the differences of the $g$- and $r$-band magnitudes of bulges and disks are less than 0.2 mag. The only exception is PGC 38112, which has a significant blue core, our  $g$- and $r$-band magnitudes of bulges are $\sim$ 1 mag brighter than those of Meert's catalog, while the magnitudes of disk (both $g$- and $r$- band) are similar. We show the decomposition result of $r$-band image in the bottom panel of Fig. 7.

In Fig 8, we show $g$-$r$ color distributions of bulge and disk for a sample of 112 S0 galaxies with the bulge and disk decomposition, containing  45 SFS0s, 5 Composites, 14 LINERs, and 48 Others, the green contour represents 670,722 SDSS DR7 spectroscopically selected galaxies from \citet{Meert2015}.
The black solid line represents galaxies having bulge and disk with equal colors. The two dashed lines are defined as $\mid (g-r)_{bulge} - (g-r)_{disk} \mid = 0.2\  mag$.
The main results from Fig. 8 are the following: 
(1) the majority of normal S0s have redder bulges than disks, the K-S statistic probability of bulge and disk from the same parent population is 0.00048,  which is consistent with whole SDSS galaxies; 
(2) 24 SFS0s  locate near the 1:1 line (black solid line), 16 SFS0s have bluer bulges than disks ($g$-$r$ bluer than 0.2 mag), and 5 ones have bluer disks than bulges, suggesting that about 1/3 SFS0s show  significant star forming activities in their bulges; 
(3) among the other 67 S0s (except for SFS0s), 43 galaxies are near to the 1:1 line, 7 objects have bluer bulges than disks, and 17 ones show bluer disks than bulges; 
(4) SFS0s have bluer bulges and disks than other types of S0s. 


In Fig. 9, we show that the S\'{e}rsic index distribution (left panel: {\em g}-band, right panel: {\em r}-band) for bulges of 45 SFS0s and 48 Others. The black color is for SFS0s, and the red one for Others S0s. We find that the median S\'{e}rsic index values of bulges for {\em g}-band and {\em r}-band for SFS0s are 1.49 and 1.68, and  3.09 and 3.46 for Others S0s, respectively. The median values of disks for {\em g}-band and {\em r}-band for SFS0s are both 1.08. It is interesting that for Others S0s, the S\'{e}rsic indices (both {\em g}-band and {\em r}-band) are similar, the majority of which are in the range of 2 to 6 ; while for SFS0s, the S\'{e}rsic indices are mainly less than 2, which might suggest that for SFS0s, the bulges are similar to disky pseudobulges.

Summarizing, through the bulge/disk decomposition, 
\textbf{we find that $\sim$ 1/3  SFS0s have bluer bulges than disks.} It is well-known that the phenomenon of gas and star misalignment is ubiquitous in S0 galaxies. \citet{Katkov2014} studied a sample of isolated S0 galaxies with long-slit spectroscopy at the Russian 6-m telescope. They find that seven of these 12 galaxies show extended ionized-gas discs whose rotation is decoupled from the stellar kinematics and five of these seven galaxies are gas-star counter-rotators. Very recently, Based on the IFU observations from MaNGA (Mapping Nearby Galaxies at APO) survey, Chen et al. (2016, submitted) find 9 star forming counter rotators, all these 9 galaxies have younger stellar population, more intense ongoing star formation in the central region than their outer disks as well as higher central metallicity, indicating ongoing bulge growth. They proposed a simple scenario to explain all the observations -- the progenitor accretes counter-rotating gas from a gas-rich dwarf or cosmological cold-gas filaments, the cancellation of angular momentum from gas-gas collisions between the pre-existing and the accreted gas largely accelerates gas inflow, leading to a burst of final star formation that is short-lived, centrally-concentrated. The higher metallicity is due to the enrichment from the final burst. Accreting counter rotating gas from external by star forming galaxies could be a new formation mechanism of S0 galaxies. 

Our SFS0 galaxies totally support this picture in the sense that (1) we find  16/45 SFS0 galaxies have obvious bluer bulges than their disks, while 5/45 SFS0 galaxies show redder bulges; (2) Fig. 10 shows the stellar mass-metallicity relation of SFS0s. The solid line represents the relation of stellar mass and metallicity for local star forming galaxies \citep{Tremonti2004}, with the two dashed lines show the $\pm$1$\sigma$ scatter. The squares are SFS0s matched with MPA/JHU catalog to get the metallicity. We find that only 3 SFS0s locate below the solid line, about half (22/45) of SFS0s show significant higher metallicities than the local stellar-metallicity relation (above the 1$\sigma$ dashed line, Tremonti et al. 2004), which is totally consistent with the properties of star forming counter rotators (Chen et al. 2016 submitted to Nature communications,  Figure 9 Jin et al. 2016, accepted by MNRAS).

\subsection{Environmental Dependence}
For studying the relationship of galaxy activity and their environment, there are a variety of methods to measure the local background density of a galaxy \citep{Muldrew2012}. The most common way is to measure the density within the distance to the Nth nearest neighbour. Following \citet{Balogh2004}, we use the projected distance, $d_5$ (the distance to its fifth nearest neighbour galaxy), to estimate a projected density ($\Sigma_5$) for each galaxy. The projected density is given by

\begin{equation}
   \Sigma_{5}=\frac{5}{\pi d_{5}^2}.
\end{equation}

\noindent In order to remove foreground and background galaxies, we limit the velocity difference of all neighbours in our sample galaxies within $\pm1000$ km s$^{-1}$. Furthermore, to avoid Malmquist bias and reach the minimum SDSS fiber separation, when we estimate $d_5$, we only consider galaxies, which are brighter than $M_{r}=-20$ mag, with separation 50 $h^{-1}$ kpc \citep{Licheng2009}.


Fig. 11 shows cumulative distribution of $\Sigma_5$ for different types of S0s. The black, orange, blue, green, and red points are SFS0s, Composite, Seyfert, LINERs, and Others, respectively. We find that Seyferts are in the most sparse environment, and SFS0s and LINERs are hardly distinguishable. In the central part of the distribution (between $\Sigma_5$ $\sim$ 0 and $\Sigma_5$ $\sim$ 1), the Composite S0s are also similar to SFS0s and LINERs. Composites and Others agree in a relatively dense environment ($\Sigma_5$ larger than 1). The median $\Sigma_5$ values of SFS0s, Composites, Seyferts, LINERs and Others are 0.723, 0.742, 0.740, 0.711 and 0.936, respectively. Since there are only 13 Seyfert S0s, the mean value of $\Sigma_5$ for Seyferts might be more reliable, which is 0.614. Our results from cumulative distribution are consistent with \citet{cv06} , where they found that LINERs are located in denser environment than Seyferts. We calculate values of K-S statistics probability of $\Sigma_5$ for SFS0s compared to Composite, Seyfert, LINERs, and Others, which are 0.55, 0.64, 0.99, and 0.37, respectively. The distributions of $\Sigma_5$ for different types of S0s are shown in Fig. 12. The histogram with black filled with grey shows the distribution for SFS0s, and orange, blue, green, and red show the distribution of $\Sigma_5$ for Composites, Seyferts, LINERs, and Others, respectively. The mean value for Seyferts and median values for SFS0s, Composites, LINERs, and Others are shown by vertical lines. Although the histograms of different types of S0s  show the similar distribution, we find that Seyfert S0s (blue dot-dashed vertical line) show the smallest value of $\Sigma_5$, and Others (red dot-dashed line) is clearly larger than other S0s, which are consistent with our previous results.


We also try another method to study the environment of galaxies for comparison, which is called the projected redshift-space two-point cross-correlation function (2PCCF), $w_p(r_p)$. In order to study the local galaxy density for our S0 galaxies, we first construct a photometric reference sample to count the companions around our S0s. The reference sample from NYU-VAGC with {\em r}-band Petrosian magnitude lower than 21 magnitude includes about 26 million galaxies. We construct 100 random samples which have the same sky coverage, magnitude, mass, and redshift limits as the reference sample (more details in Li et al., 2006a,b). We then count the number of companions in reference galaxies and random samples for each object in our S0 samples within the projected radius $r_{\rm p}$ and within $\pi$ which is the separation parallel to the line of sight.
  Here $\xi(r_p,\pi)$ is calculated using the estimator
\begin{equation}
\xi(r_p,\pi) = \frac{N_R}{N_D} \frac{QD(r_p,\pi)}{QR(r_p,\pi)} -1,
\end{equation}
where $N_D$ and $N_R$ are the number of reference sample and random  sample, $r_p$  and $\pi$ are the separations perpendicular and parallel to the line of sight, and $QD(r_p,\pi)$  and $QR(r_p,\pi)$ are the number of S0s-reference sample and S0s-random sample pairs, respectively. Following a standard approach, we focus on the projection of  $\xi(r_p,\pi)$ along the line of sight ($\pi$-direction): \begin{equation}
w_p(r_p)=\int_{-\infty}^{+\infty}\xi(r_p,\pi)d\pi=
\sum_i\xi(r_p,\pi_i)\Delta\pi_i.
\end{equation}

In Fig. 13, we compare projected redshift-space 2PCCF $w_p(r_p)$ for our five subsamples of S0s. The black, orange, blue, green, and red are SFS0s, Composite, Seyfert, LINERs, and Others, respectively. SFS0s seem more similar to Seyferts, while LINERs are quite different and Composites are similar to Others. Although the error bars are relatively large, there is still a tendency for S0s classified as Others to be in the densest environment, which is consistent with the projected density method.

It is worth comparing our results with other studies concerning the relations between S0s classified as Others and the local environment. Our result is within our expectation. Since the S0s classified as Others are quiescent galaxies or galaxies with low-level star formation or AGN activities, they have red colors and bulge-dominated morphologies, and tend to have older stellar population. The morphology-density relation of \citet{Dressler1980} showed that spheroidal systems reside preferentially in dense regions. \citet{Licheng2006a} demonstrated that the redder galaxies of all luminosities cluster more strongly than their blue counterparts. Galaxies with red colors, old stellar populations and bulge-dominated morphologies reside preferentially in the dense environment \citep{Zehavi2005}. \citet{Hogg2003} and \citet{Yang2005} found that the bright early-type galaxies are more centrally concentrated than the fainter and late-type galaxies. These findings suggest that the high frequency of interaction in dense environment can significantly quench the star formation in the galaxies.

\section[]{Conclusions}
We constructed a sample of nearby S0 galaxies by matching the SDSS DR7 and RC3 catalog. We then studied the relationship between S0s activity and environment in the local universe. By using the bulge/disk decomposition, we also discussed the properties of disks and bulges for S0s. Our main purpose is to investigate the nuclear activities of nearby S0s.

The main results of our work are as follows:

\begin{enumerate}

 \item From a sample of 583 local ($z<0.1$) S0s, we discover 232 active galaxies from BPT diagnostic diagram, and 45 galaxies out of 583 S0s with star-forming activity. The fraction of these SFS0s and active S0s are about 8 \% and 40 \%, respectively. Among the 45 SFS0s, we derive the mean nuclear SFR is 0.43 $\pm$ 0.71 $M_{\odot}$ $yr^{-1}$.

 \item According to the distribution of general properties for different S0s, SFS0s show the expected differences of galaxy fundamental properties from other S0 galaxies: lower mass, lower redshift, and lower luminosities. Because of star formation, these SFS0s are bluer than other types of S0s. Based on the above findings, we can explain the star formation activity in these low-mass systems by means of the `down-sizing' effect. 

 \item After performing 2D bulge-disk decomposition of 45 SFS0s with \textsc{Galfit} software and using \citet{Meert2015} two-dimensional photometric decompositions catalog, we get 112 S0 galaxies with bulge and disk properties.
With the distribution of bulge and disk components of S0 galaxies, we find the majority of S0s have redder bulges than disks except SFS0s. Only five of 45 SFS0s show redder bulges, and bulges and disks of 45 SFS0s are bluer than other S0s. The S\'{e}rsic index of the bulge for most our SFS0 sample is less than 2, suggesting that bulges of SFS0s might be pseudobulges; while in Others, the bulge S\'{e}rsic index is between 2 and 6, though our sample size is too small to derive the robust conclusion. We propose the formation of S0s may be due to accrete counter rotating gas from external by star forming galaxies.

 \item To study the environmental dependence of S0s activity in the local universe, we plot the cumulative distribution of $\Sigma_5$ for different types of S0s. We find that the active S0s/Others locate mainly in sparse/dense environment. From sparse to dense environment, most of S0s in sparse environment are Seyferts, the composite S0s are similar to SFS0s and LINERs in the central part of the distribution (between about $\Sigma_5$=0 and $\Sigma_5$=1), Composites and Others are in relatively dense environment ($\Sigma_5$ larger than 1).

\end{enumerate}

\section*{Acknowledgments}
The authors are very grateful to the anonymous referee for his/her critical comments and instructive suggestions, which significantly strengthened the analyses in this work.
We also thank Prof. Cheng Li and Enci Wang for helpful suggestions about calculating the galaxy environment by method of 2PCCF, thank Dr. Alan Meert for many helpful advice of using the two-dimensional photometric decompositions catalog, and 
thank Dr. Song Huang, Yifei Jin and Hai Yu for valuable discussions and advice which improved this paper. This work was supported by the National Natural Science Foundation of China (No. 11273015 and 11133001), and National Basic Research Program (973 program No. 2013CB834905).

Funding for the Sloan Digital Sky Survey (SDSS) has
been provided by the Alfred P. Sloan Foundation, the Participating
Institutions, the National Aeronautics and Space Administration, the
National Science Foundation, the US Department of Energy, the
Japanese Monbukagakusho, and the Max Planck Society. The SDSS Web
site is http://www.sdss.org.
The SDSS is managed by the Astrophysical Research Consortium (ARC)
for the Participating Institutions. The Participating Institutions
are The University of Chicago, Fermilab, the Institute for Advanced
Study, the Japan Participation Group, The Johns Hopkins University,
Los Alamos National Laboratory, the Max- Planck-Institute for
Astronomy (MPIA), the Max-Planck-Institute for Astrophysics (MPA),
New Mexico State University, University of Pittsburgh, Princeton
University, the United States Naval Observatory and the University
of Washington.

\newpage
\begin{figure}[hbt]
  \centering
  \includegraphics[width=10cm]{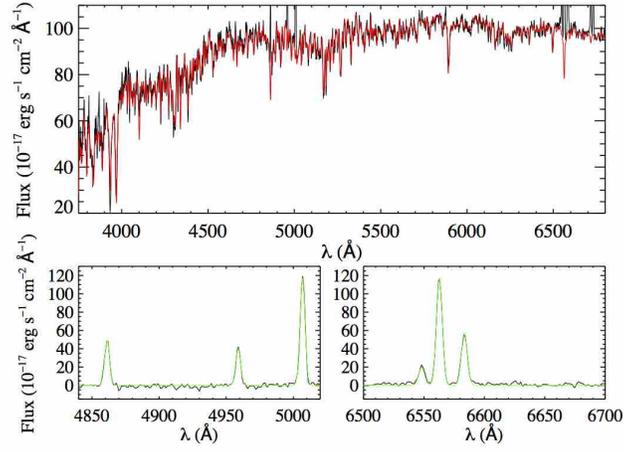}\\
  \caption{Example of spectral fitting results for PGC 33459. The observed spectrum is in black and the best fit model is in red. Bottom panels show H$\beta$ and [{O~\sc iii}]$\lambda\lambda$4959,5007 (left), H$\alpha$, [{N~\sc ii}]$\lambda\lambda$6548,6584 (right) emission line regions. The emission lines fitted by one Gaussian component each are in green, while the pure emission line spectra are in black.}\label{fitfig}
\end{figure}

 \begin{deluxetable}{lccccccc}\label{properties}
\tablecolumns{8} \tablewidth{0pc} \tabletypesize{\scriptsize}
\tablecaption{The physical properties of our sample of S0 galaxies.}
\tablehead{ \colhead{Classification} & \colhead{Number} & \colhead{$M_\ast$} & \colhead{$\sigma_\ast$} & \colhead{area} & \colhead{$z$} & \colhead{$M_r$} & \colhead{$g-r$} \\
\colhead{} & \colhead{} & \colhead{$\log(M_\odot)$} & \colhead{km/s} & \colhead{$kpc^2$} & \colhead{} & \colhead{mag} & \colhead{} } \startdata

Star-forming & 45 & $9.39 (_{-0.99} ^{+1.01})$ & $88.67 (_{-10.37} ^{+33.48})$ & 0.17 & 0.0075 & -18.30 & 0.54 \\ 
Composite & 49 & $10.62 (_{-0.60} ^{+0.56})$ & $172.94 (_{-52.44} ^{+72.77})$ & 1.26 & 0.0201 & -20.56 &0.75 \\ 
Seyfert & 13 & $10.88 (_{-1.54} ^{+0.51})$ & $184.80 (_{-86.22} ^{+65.76})$ & 1.52 & 0.0220 & -20.73 & 0.72 \\ 
LINERs & 125 & $10.92 (_{-0.41} ^{+0.31})$ & $203.23 (_{-51.49} ^{+44.24})$ & 1.60 & 0.0225 & -21.07 & 0.77 \\ 
Others & 351 & $10.85 (_{-0.50} ^{+0.37})$ & $201.66 (_{-56.62} ^{+57.80})$ & 1.79 & 0.0238 & -21.11 & 0.76 \\
\enddata
\\
\end{deluxetable}

\begin{deluxetable}{lcccc}\label{coefficient}
\tablecolumns{5} \tablewidth{0pc} \tabletypesize{\scriptsize}
\tablecaption{The correlation coefficients and the p-values of the relations between the physical properties of our S0 galaxies.}

\tablehead{\colhead{Classification} & \multicolumn{4}{c}{correlation coefficient/p-value}	\\
\colhead{} & \colhead{$M_\ast$ vs. $\sigma_\ast$} & \colhead{$M_\ast$ vs. $M_r$} & \colhead{$M_\ast$ vs. $g-r$} & \colhead{$M_r$ vs. $g-r$}} \startdata

Star-forming & 0.71/$1.0 \times 10^{-7}$ & -0.92/$1.0 \times 10^{-7}$ & 0.65/$1.43 \times 10^{-6}$ & -0.40/$6.79 \times 10^{-3}$ \\
Composite & 0.63/$5.48 \times 10^{-6}$ & -0.70/$3.58 \times 10^{-7}$ & 0.53/$2.28 \times 10^{-4}$ & -0.29/ $5.75 \times 10^{-2}$ \\
Seyfert & 0.90/$3.28 \times 10^{-5}$ & -0.88/$8.96 \times 10^{-5}$ & 0.47/0.11 & -0.37/0.21 \\
LINERs & 0.85/$1.0 \times 10^{-7}$ & -0.98/$1.0 \times 10^{-7}$ & 0.42/$1.91 \times 10^{-6}$ & -0.30/$8.10 \times 10^{-4}$ \\
Others & 0.87/$1.0 \times 10^{-7}$ & -0.98/$1.0 \times 10^{-7}$ & 0.54/$1.0 \times 10^{-7}$ & -0.47/$1.0 \times 10^{-7}$ \\
\enddata
\\
\end{deluxetable}

 \begin{deluxetable}{lccc}\label{probability}
\tablecolumns{4} \tablewidth{0pc} \tabletypesize{\scriptsize}
\tablecaption{Values of K-S statistic probability for SFS0s compared to Composite, Seyfert, LINERs, and Others.}
\tablehead{ \colhead{Classification} & \colhead{$M_\ast$} & \colhead{$M_r$} & \colhead{$g-r$} \\
\colhead{} & \colhead{$\log(M_\odot)$} & \colhead{mag} & \colhead{} } \startdata

Composite & $1.96 \times 10^{-6}$ & $2.89 \times 10^{-7}$ & $6.05 \times 10^{-11}$ \\ 
Seyfert & $2.54 \times 10^{-4}$ & $2.54 \times 10^{-4}$ & $1.41 \times 10^{-4}$ \\ 
LINERs & $4.39 \times 10^{-18}$ & $3.60 \times 10^{-16}$ & $3.31 \times 10^{-18}$ \\ 
Others & $1.03 \times 10^{-18}$ & $1.67 \times 10^{-17}$ & $6.84 \times 10^{-19}$ \\
\enddata
\\
\end{deluxetable}

\begin{figure}[hbt]
  \centering
  \includegraphics[width=12cm]{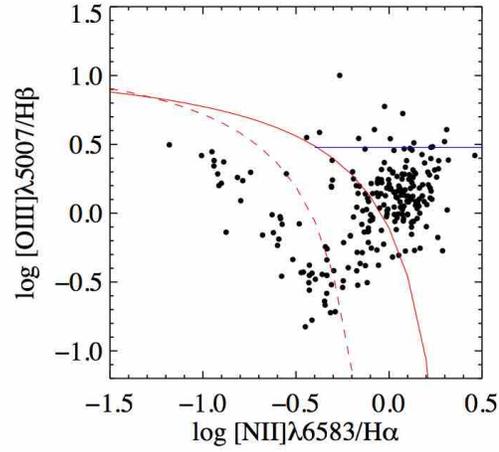}\\
  \caption{The BPT diagram of S0 galaxies. The red solid line is from the criterion of \citet{Kewley2001}, the red dash line and blue line are the criteria from \citet{Kauffmann2003a}. In this diagram, star-forming galaxies are below the red-dashed line, composite galaxies between the red-dashed and the solid line, Seyfert galaxies are in the top-right region, and LINERs are in the bottom-right region.}\label{bptfig}
\end{figure}

\begin{subfigures}
\begin{figure}[hbt]
  \centering
  \includegraphics[width=10cm]{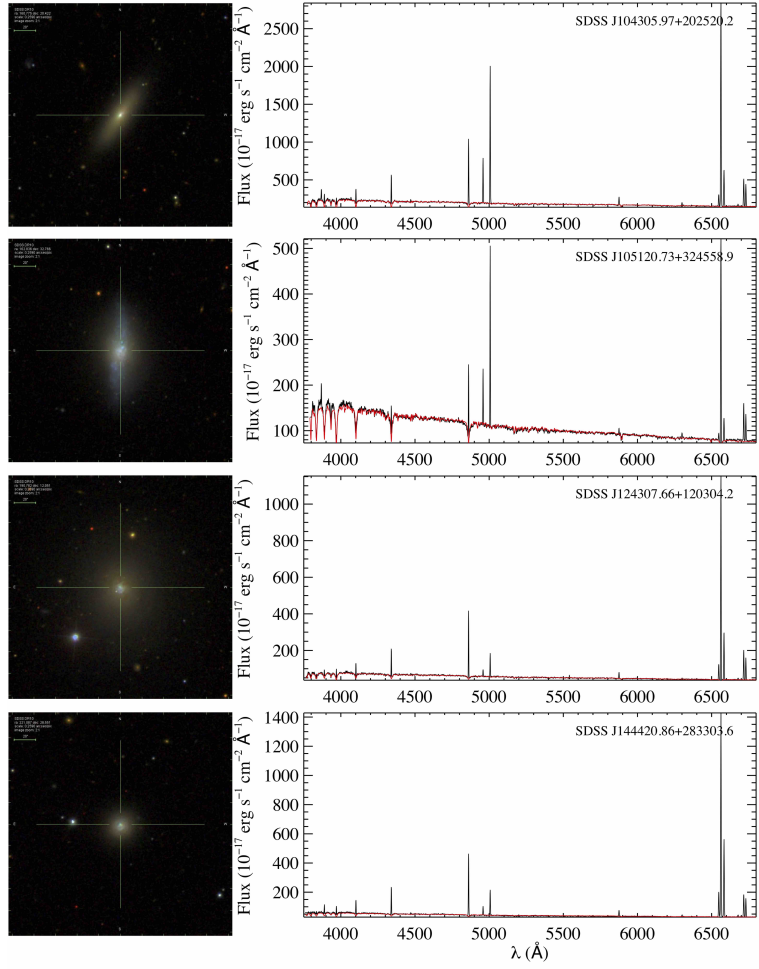}\\
  \caption{The false color({\em g}-, {\em r}- and {\em i}-bands) images and fitted spectra of Star-forming S0 galaxies. The black lines are observed spectra and the red ones are the best fitting spectra. }\label{SFs_fig}
\end{figure}

\begin{figure}[hbt]
  \centering
  \includegraphics[width=10cm]{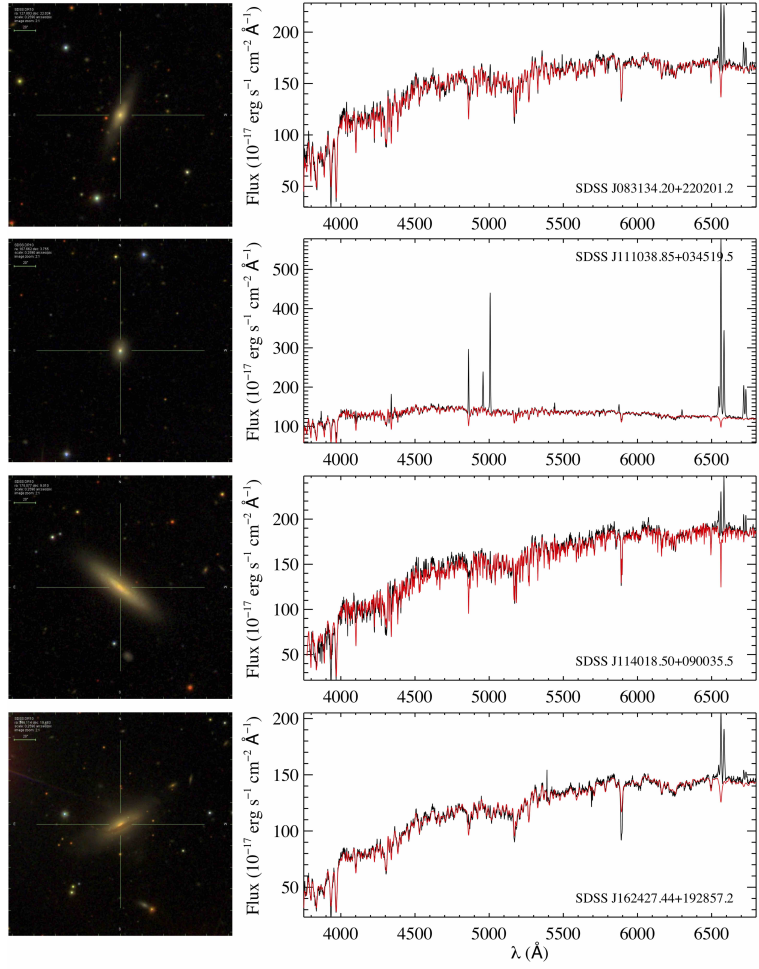}\\
  \caption{The same as in Fig. 3a for Composite S0 galaxies.}\label{composite_fig}
\end{figure}

\begin{figure}[hbt]
  \centering
  \includegraphics[width=10cm]{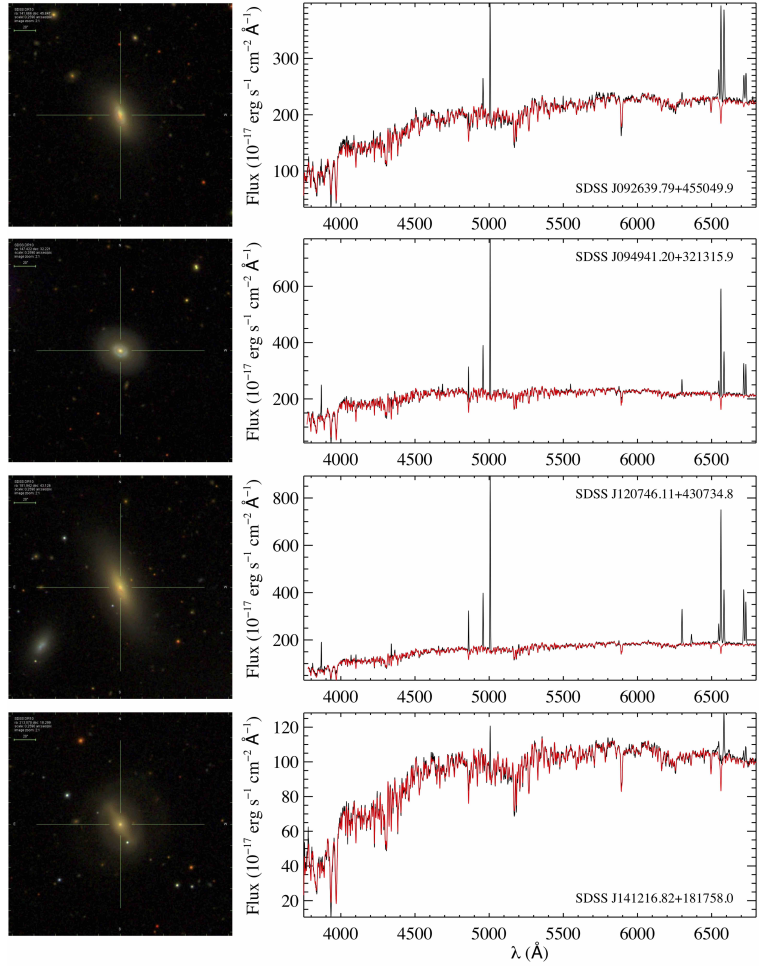}\\
  \caption{The same as in Fig. 3a for Seyfert S0 galaxies.}\label{seyfert_fig}
\end{figure}

\begin{figure}[hbt]
  \centering
  \includegraphics[width=10cm]{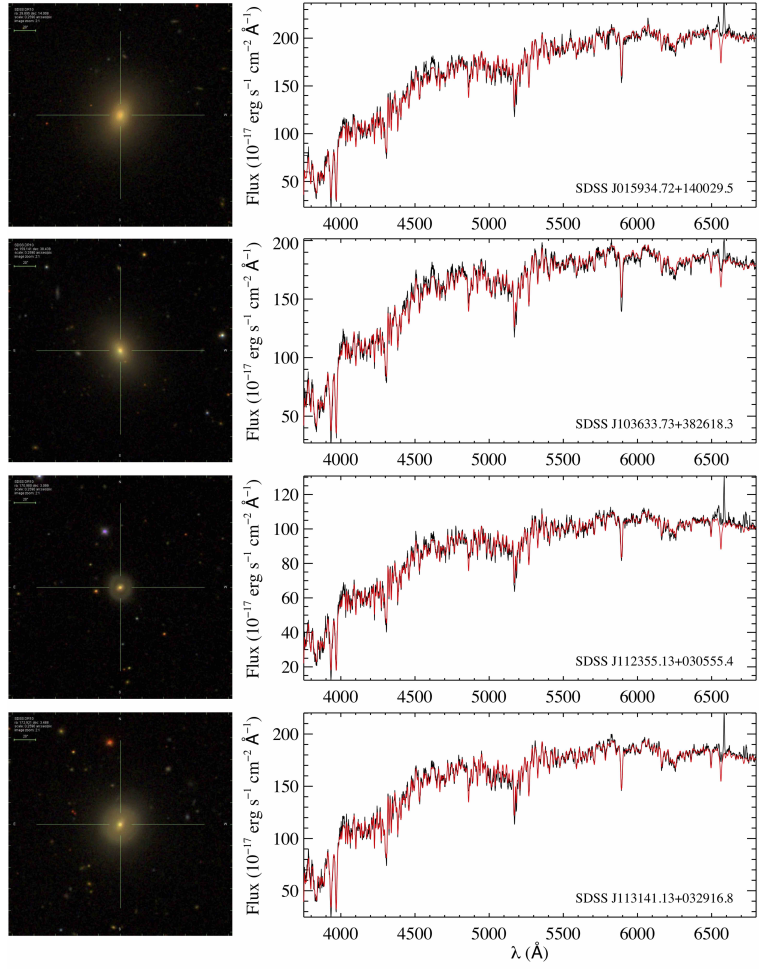}\\
  \caption{The same as in Fig. 3a for LINERs S0 galaxies.}\label{liners_fig}
\end{figure}

\begin{figure}[hbt]
  \centering
  \includegraphics[width=10cm]{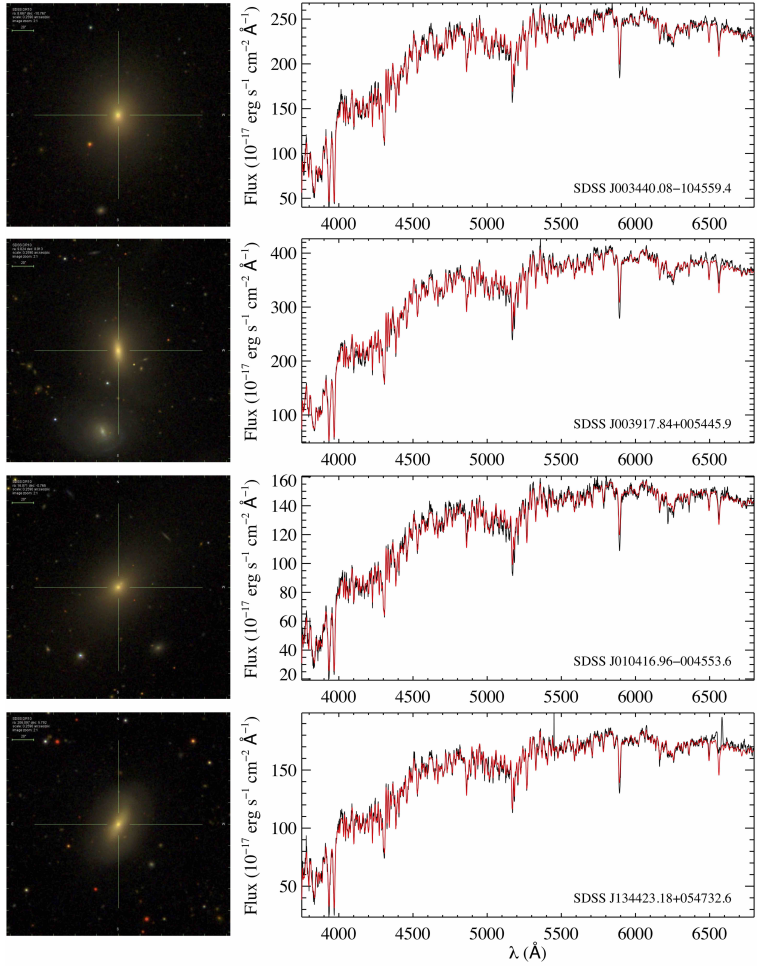}\\
  \caption{The same as in Fig. 3a for Others S0 galaxies.}\label{others_fig}
\end{figure}
\end{subfigures}

\begin{figure}[hbt]
  \centering
  \includegraphics[width=10cm]{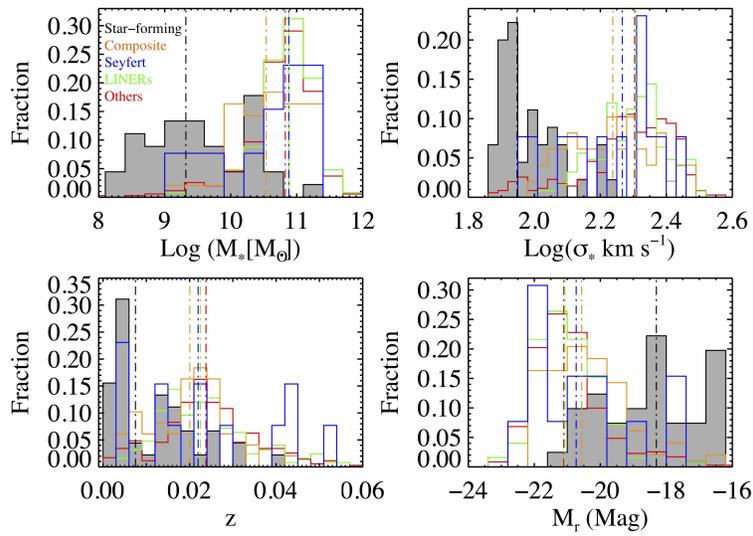}\\
  \caption{Distribution of stellar mass (top-left), velocity dispersion (top-right), redshift (bottom-left) and {\em r}-band absolute magnitude (bottom-right) of our different types of S0s. The fraction represents normalized values of numbers for different types of S0s, separately. The vertical lines show median values for different types of S0s. In these histograms, Star-forming, Composite, Seyfert, LINERs, and Others are shown in black (filled with grey), orange, blue, green, and red, respectively.}\label{allmassfig}
\end{figure}

\begin{figure}[hbt]
  \centering
  \includegraphics[width=10cm]{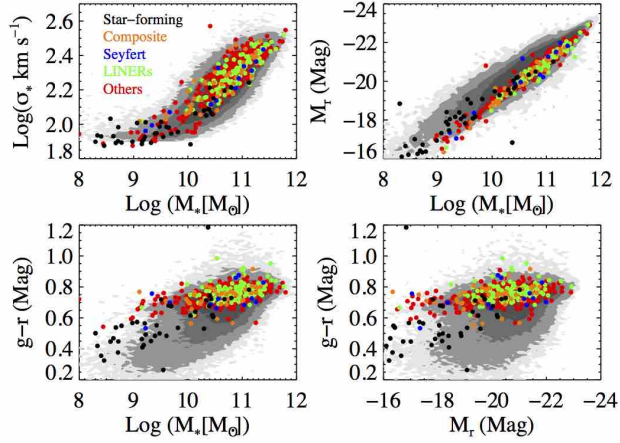}\\
  \caption{The relation between stellar mass, velocity dispersion, {\em r}-band absolute magnitude and $g-r$ color. In this diagram, SFS0s are black circles, Composite S0s are orange, Seyferts are blue, LINERs are green and Others are red. The background grey contours represent all galaxies with spectral information from SDSS DR7 \citep{Abazajian2009}, we limit the median S/N to values larger than 20.}\label{combine06fig} 
\end{figure}

\begin{figure}[hbt]
  \centering
  \includegraphics[width=10cm]{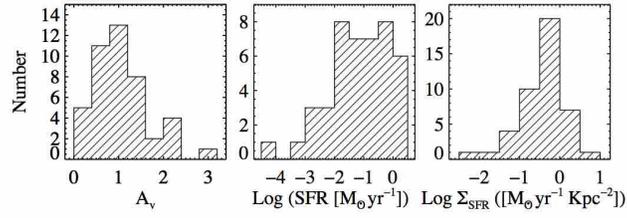}\\
  \caption{The histograms of nebular extinction, SFRs, and SFRs per unit area for 45 star-forming S0 galaxies.}\label{sfrfig}
\end{figure}

\newpage
\begin{figure}[hbt]
  \centering
  \includegraphics[width=10cm]{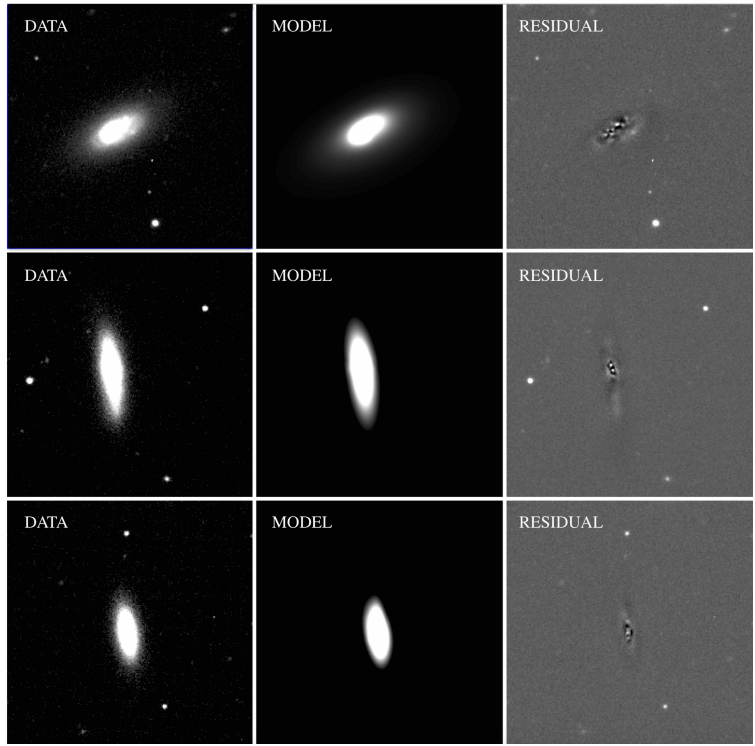}\\
  \caption{Three examples of the bulge and disk decomposition with \textsc{Galfit}. The top three panels are PGC 36750, where from left to right, the {\em g}- band observed image, model image, and model-subtracted residual image are displayed. The middle and the bottom three panels are the {\em r}- band results for PGC 49734 and PGC 38112. The foreground stars have been masked before fitting. }\label{decompositionfig}
\end{figure}

\newpage
\begin{figure}[hbt]
  \centering
  \includegraphics[width=10cm]{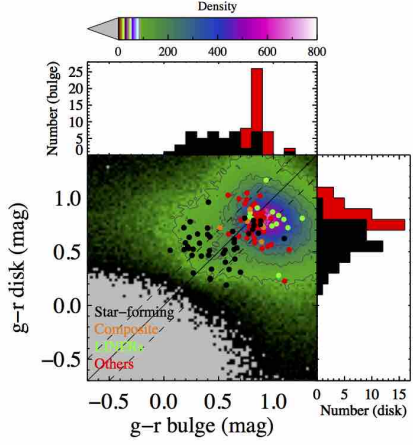}\\
  \caption{The $g$-$r$ color distribution of bulges and disks for S0 galaxies. The black, orange, green, and red points represent SFS0s, Composite, LINERs, and Others, respectively. The background green contours are data from the two-dimensional decomposition catalog of  Meert et al. (2015). The black solid line represents the galaxies whose bulges and disks have the same $g-r$ colors. The two dashed lines are defined as $\mid (g-r)_{bulge} - (g-r)_{disk} \mid = 0.2$. The histograms of $g-r$ colors for bulges and disks are in the upper panel and in the right panel, respectively. The top panel is a color bar of the main panel.}\label{his_4_config}
\end{figure}

\newpage
\begin{figure}[hbt]
  \centering
  \includegraphics[width=10cm]{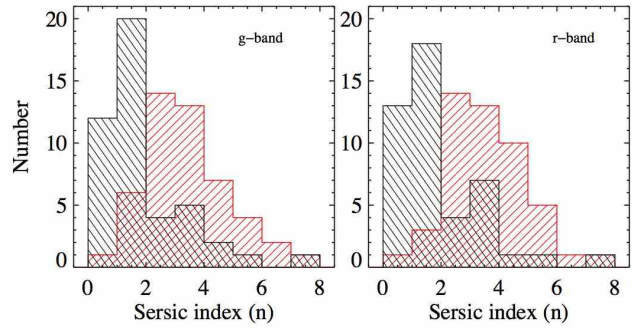}\\
 \caption{Bulge S\'{e}rsic index distributions for SFS0s (black right slash) and Others (red left slash) with {\em g}-band and {\em r}-band.}\label{n_bulgefig}
\end{figure}

\begin{figure}[hbt]
  \centering
 \includegraphics[width=10cm]{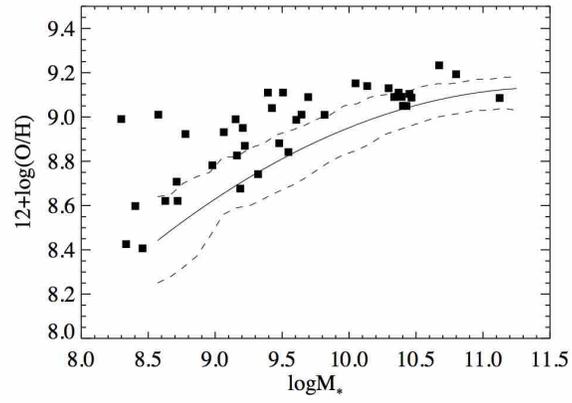}\\
  \caption{The stellar mass-metallicity relation of SFS0s. The solid line shows the relation for local star forming galaxies given by \citet{Tremonti2004}. The two dashed lines are the $\pm$1$\sigma$ scatter region.}\label{m_Zfig}
\end{figure}

\newpage
\begin{figure}[hbt]
  \centering
  \includegraphics[width=10cm]{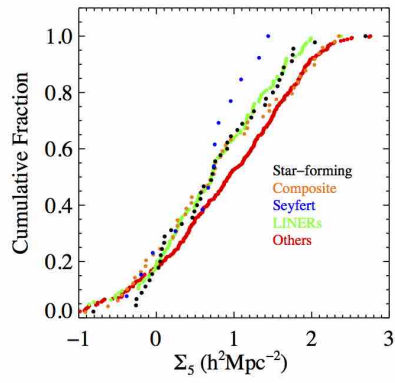}\\
  \caption{The cumulative fraction of projected density ($\Sigma_5$) for different types of S0 galaxies.}\label{cumulate1fig}
\end{figure}

\newpage
\begin{figure}[hbt]
  \centering
  \includegraphics[width=10cm]{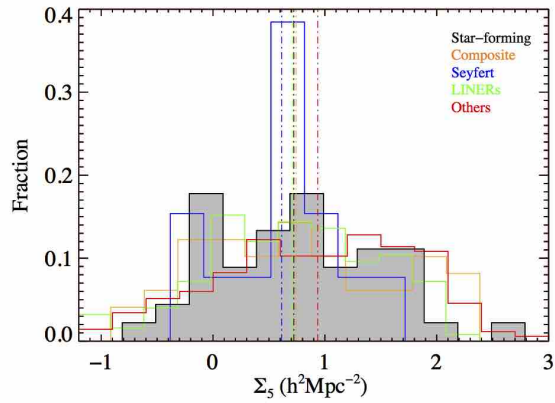}\\
  \caption{Distribution of projected density ($\Sigma_5$) for SFS0s (black filled with grey), Composites (orange), Seyferts (blue), LINERs (green) and Others (red). The fraction represents normalized values of numbers for different types of S0s, separately. The vertical lines show mean value for Seyferts and median values for SFS0s (black), Composites, LINERs, and  Others.}\label{dens_fig}
\end{figure}

\newpage
\begin{figure}[hbt]
  \centering
  \includegraphics[width=10cm]{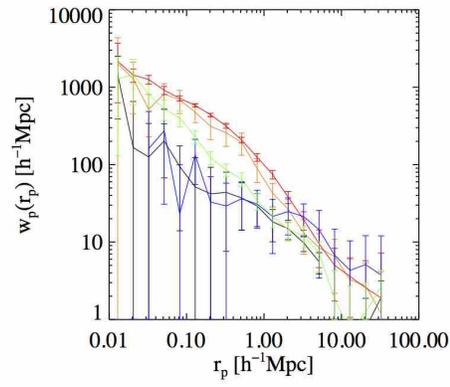}\\
  \caption{Projected redshift-space 2PCCF $w_p(r_p)$ for SFS0s (black), Composites (orange), Seyferts (blue), LINERs (green) and Others (red).}\label{enviro_errfig}
\end{figure}

\end{document}